# A metastable amorphous intermediate is responsible for laser-induced nucleation of glycine

Zhiyu Liao, Klaas Wynne*

School of Chemistry, University of Glasgow, G12 8QQ, UK

ABSTRACT: Laser-induced crystal nucleation through optical tweezing, and in particular polymorph selection through laser polarization, promises unprecedented control over crystallization. However, in the absence of a nearby liquid–liquid critical point or miscibility gap, the origin of the required mesoscale clusters remains unclear. A number of recent studies of so-called nonclassical nucleation have suggested the presence of large amorphous clusters. Here we show that supersaturated aqueous glycine solutions form metastable intermediate particles that are off the direct path to crystal nucleation. Laser-induced crystal nucleation only occurs when the laser "activates" one of these particles. *In situ* low-frequency Raman spectroscopy is used to demonstrate their amorphous or partially ordered character and transformation to various crystal polymorphs. The requirement for solution aging in many previously reported laser-induced crystal nucleation experiments strongly suggests that the presence of amorphous intermediates is a general requirement.

## INTRODUCTION

Nucleation of crystals from solution is of great importance—ranging from the production of pharmaceutical drugs and chemical products to biological crystallography—but is difficult to control. In the mid-1990s, it was discovered that nucleation can be greatly accelerated by exposing supersaturated solutions to nanosecond pulsed-laser radiation[1] and that the crystal polymorph could be selected by using either linearly or circularly polarized laser light.[2] This phenomenon was referred to as non-photochemical laser-induced nucleation (NPLIN) to emphasize that the nucleation process does not involve the making or breaking of any chemical bonds. Subsequently, several groups demonstrated that pulsed lasers may be used to induce nucleation in supersaturated solutions,[3-6] although the polymorph selection ability has been questioned.[7] More recently, it has been shown that continuous-wave (CW) lasers can also induce crystal nucleation at the solution–air interface.[8]

Initially it was suggested that the optical Kerr effect might be responsible for NPLIN. In the optical Kerr effect, a laser induces a dipole moment in anisotropically polarizable molecules, giving rise to a force that tends to align the molecules with the polarization direction of the laser, which it was suggested could lower the barrier to nucleation. However, this effect is very weak, and it was shown through Monte Carlo simulations that at reasonable concentrations at which the solute molecules are individually dispersed, significantly higher laser powers than those used in the experiments are required.[9]

The induction of nucleation by CW lasers was explained by the effect of optical trapping. In standard applications of optical trapping, a tightly focused laser beam exerts a force holding a particle (often a glass bead) in place if the particle has a refractive index larger than that of the surrounding medium.[10] It was suggested that that this effect could trap clusters of solute molecules and hence again lower the barrier to nucleation. However, the forces associated with optical trapping of sub-micron objects are expected to be insignificant compared to thermal fluctuations induced by laser heating. Moreover, unless the system is close to a liquid–liquid phase transition,[11-15] there is no obvious physical reason for the presence of such molecular clusters. Therefore, the mechanism behind laser-induced crystal nucleation far away from any liquid–liquid critical points or in the absence of a liquid–liquid miscibility gap is still a mystery.

Could other supramolecular structures be involved in these laser-induced crystal nucleation phenomena? Previous work on tartaric acid solutions has indicated the presence of unidentified clusters of an intermediate form just before a crystal nucleates,[16] while $CaCO_3$ is thought to crystallize by way of a nanosized liquid-like amorphous calcium carbonate phase.[17,18] Light and small-angle x-ray scattering studies have shown the presence of ~250-nm glycine rich nanodroplets in saturated glycine solutions,[19] while large mesoscale clusters (> 700 nm) were detected by *in situ* dynamic light scattering (DLS) during laser induced phase separation (LIPS) experiments.[20] Despite the evidence of such nanometer-scale species in solution, very little about their state and structure is known.

Here, we investigate homogeneous and laser-induced crystal nucleation of glycine from solution using microscopy and broadband *in situ* Raman spectroscopy to provide detailed information on mesoscale clusters and their transformation to the crystalline form. The nucleation of glycine has been widely studied including the possible effects of stirring,[21] non-photochemical laser-induced crystal nucleation,[22,23] and nucleation through CW optical trapping including polymorph control through laser polarization.[24] We will show that supersaturated glycine solutions produce metastable amorphous particles whose relative stability impedes the spontaneous nucleation of a crystalline form. Irradiation of the intermediate particles with a laser induces rapid nucleation and formation of one of the polymorphs of glycine.

## EXPERIMENTAL

Saturated glycine/$D_2O$ solutions (S = 1.0) were prepared by dissolving 0.22 g of glycine in 1 g of $D_2O$ in a clean glass vial at room temperature (~ 21°C). To minimize heating of the sample by absorption of the near-infrared laser by vibrational overtone bands, solutions were prepared in $D_2O$. Nucleation experiments were carried out in a setup consisting of an inverted microscope allowing access by high-power lasers (either a pulsed 1040 nm laser, with a 4 ps pulse width, a 1 MHz repetition rate, and an average power of ≤8 W or a CW 1064 nm laser with a power of ≤10 W) and a single-frequency CW Raman excitation laser (532 nm, ≤500 mW). Back scattered light is collected confocally to provide Stokes and anti-Stokes Raman spectra with a minimum shift of about 10 $cm^{-1}$ (See Methods in SI and Figure S1). Raman spectroscopy was carried out with the Raman excitation laser power reduced to 50 mW. The local temperature elevation in the laser focus during these experiments is estimated to be ~3K by using anti-Stokes and Stokes Raman scattering of the glycine solution (see Methods in SI and Figure S2 for details). Control experiments were performed using dynamic light scattering to rule out the possibility of dust as particles observed in glycine/$D_2O$ solution (see Methods in SI and Figure S3 for details).

Any microscopic particles or crystals in the sample can be optically trapped by the laser beam(s). However, the sample environment chamber and temperature control necessitated the use of a relatively low numerical aperture objective (NA 0.7), which provided trapping in the xy-plane only. In this setup, the z-direction scattering force pushed the particles or crystals into the solution–air interface.

Aqueous glycine solution samples were prepared by placing 10 µL of a saturated or under-saturated solution onto the glass bottom of a 12-mm diameter polymer and glass Petri dish that was covered with a lid (see Figure 1). In laser-induced nucleation experiments, additional droplets of identical solution with a total volume of 1 mL were placed on the side to slow down evaporation. With this arrangement, no spontaneous crystallization was observed over a period of ≥18 hours. For homogeneous nucleation experiments, the same setup was employed, except that the additional solution droplets were replaced with Drierite particles to accelerate evaporation.

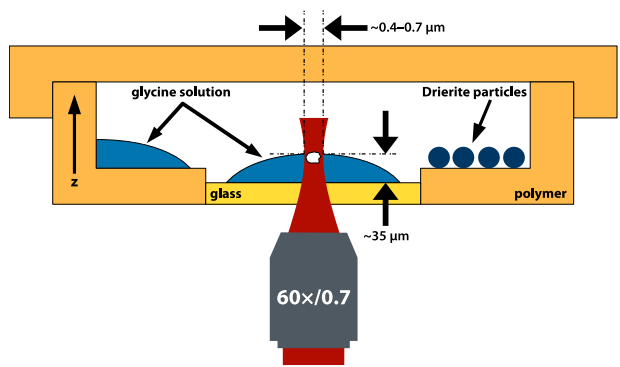

**Figure 1. Schematic of the sample setup for homogeneous nucleation and laser-induced nucleation experiments.** A glycine/$D_2O$ solution droplet was placed on the central glass portion of a plastic Petri dish to be investigated using phase contrast microscopy and confocal Raman microscopy. In laser-induced nucleation experiments, identical glycine solution droplets were placed on the side to *slow down* evaporation. In homogeneous nucleation studies, Drierite particles were placed on the side to *accelerate* evaporation.

## RESULTS

### Homogeneous nucleation

To determine the spectral features that may be expected during crystal nucleation, Raman spectroscopic investigations of homogeneous crystallization of glycine from $D_2O$ solution were performed. A droplet of 10 µL glycine/$D_2O$ solution (starting saturation *S* = 0.5) was placed in the glass Petri dish, with Drierite particles on the side, and covered with a lid. Broadband Raman spectra were recorded throughout the process with the measured power of the 532-nm excitation laser 50 mW at the sample.

Figure 2(a) shows the spectra at selected time points. By monitoring the intensity of the peak at 1410 $cm^{-1}$ (glycine $CO_2$ symmetric stretching; see left inset of Figure 2(a)), one can estimate that the glycine supersaturation quadruples due to solvent evaporation from 0.5 (1.665 M) to 2 (6.66 M) just before nucleation occurs. As the concentration increases, the double humped band peaking at ~60 $cm^{-1}$ and 170 $cm^{-1}$ (attributed to hydrogen-bond bending and stretching modes[25]) flattens out. Subtracting the spectrum of pure $D_2O$ reveals two hidden bands centered at ~70 $cm^{-1}$ and 290 $cm^{-1}$ (most likely due to glycine librations[26] and water in the solvation shell of glycine[27]) emerging over time (see right inset of Figure 2(a)).

When the nucleation of a glycine crystal occurs at 2970 s, dramatic changes can be seen in the low-frequency region (see Figure 2(b)) by the appearance and growth of several sharp peaks near ~100 $cm^{-1}$ due to lattice phonon modes of the $\beta$ polymorph.[28] From the difference spectra, it appears that the bands at 70 cm$^{-1}$ and 290 cm$^{-1}$ split into well-defined sharp peaks and upshift by about 30 cm$^{-1}$ upon nucleation. The solution was subsequently dried, and the polymorph of the crystals determined using Raman spectroscopy. It was found that $\alpha$ and $\gamma$ polymorphs had formed from the initial $\beta$ polymorph.

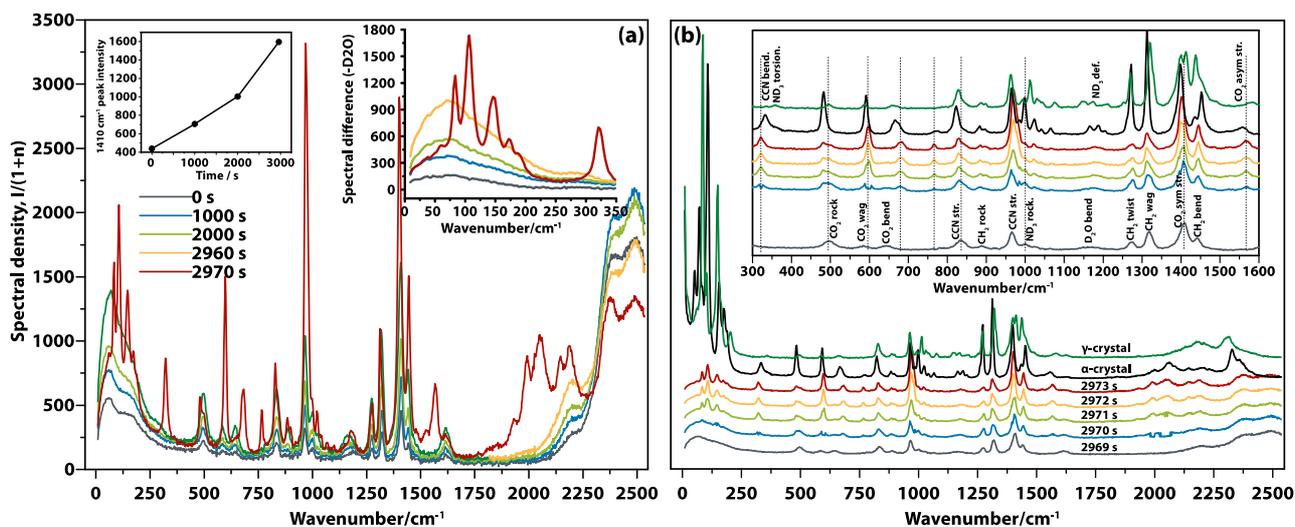

**Figure 2. Homogeneous crystal nucleation of glycine in D$_2$O solution monitored by broadband Raman spectroscopy**. (a) Each line represents the average of 10 spectra with a 1-s integration time. The inset on the left shows the 1410 cm$^{-1}$ peak intensity up to 2960 s. The inset on the right shows the spectral difference in the low frequency region after subtracting the Raman spectrum of pure D$_2$O. (b) Raman spectra of the glycine/D$_2$O solution in the few seconds before crystallization (integration time 1 s). The inset shows the spectra in the fingerprint region with peak assignments.[29]

## Nucleation from particles

To investigate laser induced nucleation of glycine crystals in solution, as suggested in the literature, the laser was focused on the air–liquid interface to initiate nucleation. Despite many attempts using ~1.0 W of either of the high-power lasers focused on the air–liquid interface for 20-30 minutes, no laser-induced nucleation was observed. However, it was observed that tiny particles (≤1 μm diameter, which is the diffraction limit in this setup) formed in the solutions. Dynamic light-scattering experiments were carried out (see Methods in SI and Figure S3) showing that these particles form on a time scale of few hours to a diameter of ~500 nm consistent with the microscopy observations. This rules out the possibility that these particles are simply dust (see next section for further evidence of Raman intensity comparison) that could potentially mediate nucleation.[30,31] Bringing one of these particles into the laser focus of one of the high-power lasers (380 mW in the sample) reliably triggered their rapid transformation into glycine crystals (faster than the 30Hz frame rate of the camera used, see Figure S4).

Figure 3 shows a typical example of a particle forming and transforming into a glycine crystal induced by high-power laser irradiation. Here particles (one of them indicated by the arrow in Figure 3(a)) formed in saturated glycine/D$_2$O solutions ($S = 1.0$) after laser irradiation for about 30 minutes. Once brought into the laser focus, the particle immediately started to crystallize in a needle-like habit and continued to grow even when the laser was blocked, demonstrating that the solution had become supersaturated. Approximately 7 needle-like crystals can be seen (Figure 3(c)), showing that the laser caused only a small number of separate nucleation events. After a few seconds, a crystal with a prismatic habit grows out on the needles and ultimately replaces the needle habit completely (see Supplementary Movie 1). Raman spectra of the two crystal habits (Figure S5) show distinct features in the low-frequency region,[28] showing that the needle-like habit is the $\beta$ polymorph and the prismatic one the $\alpha$ polymorph of glycine consistent with previous observations.[32]

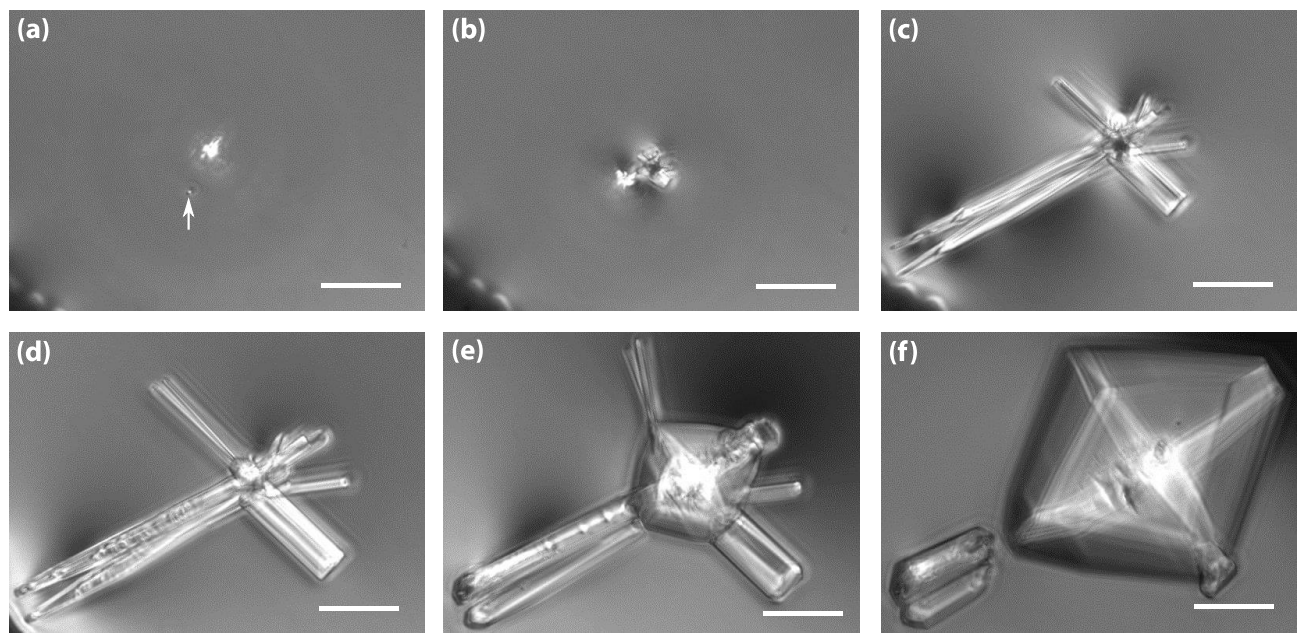

**Figure 3. Laser-induced nucleation and subsequent growth from a glycine particle in solution.** Scale bars, 30 µm. (a) One of the micrometer sized glycine particle in solution, indicated by the arrow. The bright spot is scattered light from the focus of the 1040-nm pulsed laser set to 380 mW in the sample; (b) The glycine particle is brought into laser focus by manually moving the stage instantly triggering crystallization; (c) Needle-like crystals grow rapidly while the laser has been switched off; (d) A prismatic crystal begins to grow out of the needles seconds later; (e) The needle like habit is subsumed by the prismatic crystal habit gradually; (f) The needle-like habit eventually separates and re-dissolves, and the prismatic crystal clears its surface.

## Raman spectroscopy

Individual particles in solution were investigated by confocal Raman microscopy to characterize their chemical makeup, formation, and the role they play in the process of laser-induced nucleation and crystal growth. Particles were observed in glycine/$D_2O$ solution droplets placed in the Petri dish with an initial saturation of $S = 1.0$ and either aged overnight or irradiated with 1 W of the 1040-nm laser for 30 minutes. After this preparation stage, broadband Raman spectra were taken using the 532 nm laser at a sufficiently low power (50 mW) to slow down the process of particles transforming into crystals.

An example of the development from a particle to crystal is given in Figure 4 (also see the corresponding Supplementary Movie 2 and Figure S6). For the first 16s the laser is focused near the solution–glass interface and the Raman spectrum is that of glass and aqueous glycine solution. The spectrum at $t = 16$s shows broad bands at ~500 $cm^{-1}$ and 900 $cm^{-1}$ from the glass substrate, an OD-stretch band peaking at 2,600 $cm^{-1}$, and numerous sharper peaks in the fingerprint region due to aqueous glycine. A particle was then optically trapped by the laser at $t = 22$ s. Due to their small size (≤1 µm), the Raman signal of the particles is weak and difficult to distinguish from the surrounding solution. However, as shown in Figure S7, when averaging from $t = 200$ to 209s and normalizing on the OD-stretch band, it is seen that the glycine fingerprint peaks are 1.5 times higher in magnitude showing that the particles are more glycine rich than the solution, which further confirms that the particle is not room dust.

At this much lower laser fluence compared to the experiments of Figure 3, the particle trapped in the laser focus remains (meta)stable for a considerable amount of time (up to $t = 209$s in Figure 4) and then rapidly expands to a larger mesoscale cluster, goes through a number of spectroscopically distinguishable stages, before the nucleation and growth of the final crystal. These changes in the spectrum can be observed from $t = 210$ s (most easily observed in the difference spectra shown in Figure 4(b)). In the low-frequency region, a relatively sharp peak at 105 $cm^{-1}$ (likely due to a lattice phonon mode) appears at $t = 210$ s, becomes stronger over time, peaks at $t = 217$ s (at which point significant morphological changes of the particle were observed by microscopy including an increase in overall size, see Figure 4(b)), then disappears for about 20 s, before emerging again at $t = 250$ s, and finally dominates strongly from $t = 283$ s.

A similar pattern is followed by the band at 60 $cm^{-1}$ except that it vanishes from $t = 283$ s at which point it splits into three peaks due to the lattice phonon modes of (in this case) α glycine. Due to its low intensity, the 60 $cm^{-1}$ band is not very obvious before $t = 218$ s after subtraction of $D_2O$ spectrum. However, by averaging multiple spectra, one can see a clear band at ~60 $cm^{-1}$ despite the noise. Similar but not identical behavior was observed in repeat experiments. For example, Figure S8 shows a more rapid transformation with the 60 $cm^{-1}$ band appearing in 10s and a more direct transformation to α glycine in 340s.

In the fingerprint region, peaks also become gradually more intense but fluctuate over time, *e.g.*, peaks at 483 $cm^{-1}$ corresponding to the $CO_2$ rocking and at 1410 $cm^{-1}$ corresponding to the $CO_2$ symmetric stretching modes. The latter downshifted by 10 $cm^{-1}$ at $t = 283$ s, which implies a weakening of the $CO_2$ double bond on formation of the final crystalline form. The sudden disappearance of the 2063 $cm^{-1}$ peak and appearance of 2198 $cm^{-1}$ peak at $t = 282$ s can be attributed to the polarization dependence of N–D symmetric and antisymmetric stretching vibrations in crystalline glycine (see Figure S9 for the polarized Raman spectra of glycine crystals) combined with the slow rotation of the crystal in the optical trap.

Thus, these results show that the particle remains (meta)stable up to $t = 209$ s, at which point it grows and undergoes structural transformations. The stochastic temporal changes of the Raman peaks in the low-frequency region (see Figure 4(c)) show that the particle goes through intermediate phases before finally becoming fully crystalline at $t = 283$ s. The amplitudes of the two bands in the low-frequency region (60 cm$^{-1}$ and 105 cm$^{-1}$) followed the same pattern, while that of the 483 cm$^{-1}$ band was slightly different, but they all stabilized at 283 s. The strong fingerprint Raman bands but broad phonon-like peaks hidden in the 60 cm$^{-1}$ band suggest that the solid structure was in an amorphous state before $t = 282$ s. The relatively sharp peak at 105 cm$^{-1}$ is a sign of a structure that has partial ordering.

In the nucleation experiments using the high-power lasers (*e.g.*, Figure 3), one would often see the nucleation of *β*-glycine followed by transformation to the *α* or *γ* polymorph. This is not typically seen in the Raman spectroscopy experiments likely due to the much poorer time resolution of ≥1s. Low-frequency Raman spectroscopy on randomly selected crystals after drying of the irradiated solution reveals that 38% (8 out of 21) are *γ*-glycine with the rest being *α*-glycine (see Figure S10). In contrast, crystals formed in a control sample (solution dried out naturally in the absence of laser irradiation) are all the *α* polymorph.

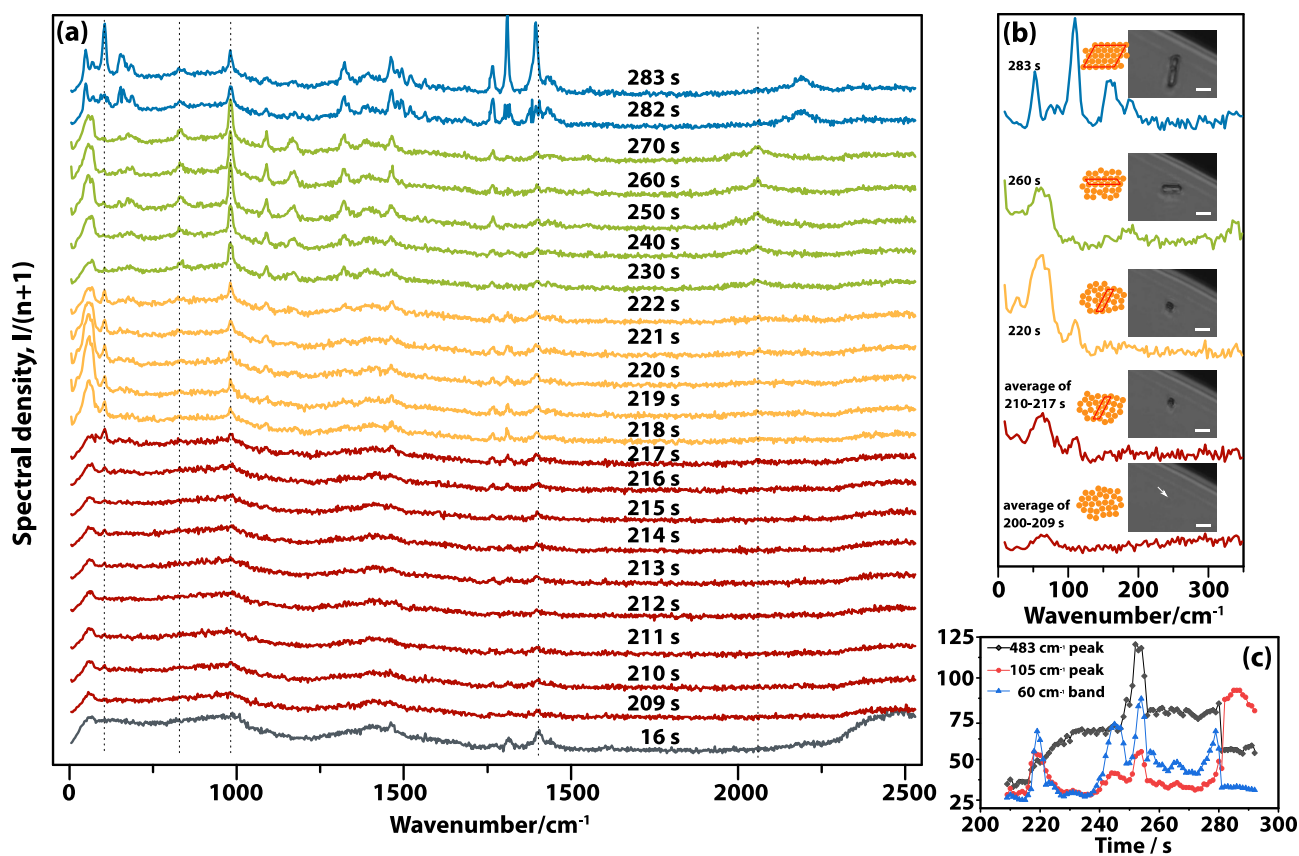

**Figure 4. Spectral evolution due to intermediate stages between microscopic particle and a fully-fledged crystal.** (a) Raman spectra taken during transitions of glycine particles in D$_2$O; (b) Selected difference Raman spectra (with the spectrum at t = 16 s subtracted) in the low-frequency region at key times, along with the time-matched microscopy images and cartoons of proposed structure of intermediates. Scale bars, 5 μm; (c) Temporal evolution of Raman peak intensities in the low-frequency region for the 60 (blue), 105 (red), and 483 cm$^{-1}$ (black) bands.

## DISCUSSION

Liquid-like clusters or droplets (hundreds of nanometers in diameter) in aqueous glycine solutions have been reported previously.[19, 33] However, the physical reason for the presence of such mesoscale clusters is still unclear. A hidden liquid–liquid critical point or miscibility gap could be the cause for concentration fluctuations or liquid–liquid phase separation leading to such clusters,[12,34] but this has not yet been found for aqueous glycine solutions.

It has been reported that aging is required for nanodroplets to form in glycine/H$_2$O solution[35] and that these nanodroplets are a requirement for laser induced nucleation to occur.[33] This is consistent with our observations: microscopic particles appear after aging or laser irradiation, and nucleation can only be induced when the particles are hit by the laser. Such a requirement for aging to improve the effectivity of laser induced nucleation has also been reported for urea,[1,7] L-histidine,[36] hen egg white lysozyme,[37] and glucose isomerase,[37] suggesting that it is a general phenomenon.

The amorphous particles we observed in solution survived for many days without exhibiting spontaneous nucleation, implying that, instead of intermediates on the path to nucleation, they appear to be off-pathway intermediates. The Raman spectroscopy experiments show that the particles have a significantly enhanced glycine concentration compared to the solution although it is difficult to determine the precise water content. A back of the envelope calculation shows that, if the particles were 100% glycine, and if the diameter is indeed ~1μm, they would contain an estimated $5\times10^9$ molecules, which is many orders of magnitude larger than typical estimates for the size of a critical nucleus (~$10^3$) in classical Gibbs theory of nucleation. This supports the idea that the particles are off path and are "activated" by the laser. This is not necessarily inconsistent with the hypothesis of a hidden liquid–liquid critical point or miscibility gap.

When the high-power lasers are used to irradiate a particle (*e.g.*, Figure 3), the laser-induced nucleation is instantaneous (≤1/30[th]s) while producing a small number of crystals (see Figure S11). When the low-power Raman excitation laser is used to irradiate a particle, the particle remains metastable for between 10 and 300s, after which a relatively slow nucleation process takes place through a number of intermediate stages, finally resulting in the formation of one of the crystal polymorphs. In 16 independent single-particle laser-induced nucleation experiments, the great majority only produced a single crystal (see Figure S12). Bands in the lattice phonon region show changes from a broad unstructured band to sharp split peaks, showing that the mesoscale cluster experienced transitions from amorphous, to partially ordered, to crystalline. This resembles the crystallization pathway of organic molecules observed previously using time-resolved cryo-TEM, where the crystalline state is formed from amorphous aggregates through a partially ordered intermediate.[38] A more recent time-resolved liquid-cell TEM study has suggested that amorphous lysozyme particles act as heterogeneous nucleation sites for lysozyme crystallization.[39] At this stage, due to the limited spatial resolution of optical microscopy, we cannot completely rule out the possibility of a similar effect in laser-induced nucleation.

The phonon–region band of the mesoscale cluster is distinct from that seen in highly concentrated solutions, showing it is a different phase. The amorphous state observed here is also different from the one detected at low temperature (T = -60°C) in previous work,[28,40] where the latter has 7 water molecules for each glycine molecule. The Raman spectrum of the mesoscale cluster (which is now easily seen in microscopy and much larger than the laser focus) has no significant OD-stretch Raman intensity (at 2392 cm$^{-1}$ and 2486 cm$^{-1}$), which implies little or no water is involved at this stage. For the same reason, it can be ruled out that the detected mesoscale cluster is glycine dihydrate.[40] Therefore, the mesoscale clusters detected here are amorphous glycine, and the glycine dihydrate phase is bypassed at room temperature in solution.

The intermediate stages between the amorphous particle and the final crystal are reminiscent of the process of prenucleation and redissolution characteristic of the classical Gibbs model of nucleation, except here in an amorphous (or partially ordered) phase within the mesoscale cluster, and under the influence of a laser field. The effect of the laser might be heating, causing reorganization within the particle that leads to a speedup of classical nucleation. On the other hand, the electric field of the laser may facilitate the crystallization from the amorphous phase by inducing net alignment and reducing the free-energy barrier through the optical Kerr effect, which may now operate on a micrometer-sized structure (providing sufficient high local concentration as opposed to individual molecules) allowing the laser polarization to influence polymorph selection. Early experiments on aqueous glycine solutions reported that the *γ* polymorph was favored in laser-induced nucleation while the *α* polymorph was always produced in spontaneous nucleation.[35] This is consistent with our finding that 38% of crystals produced in the laser-induced nucleation experiments are *γ* polymorph whereas an unirradiated sample that is dried out just forms the *α* polymorph. This supports the optical Kerr effect hypothesis.

It is generally understood that (and indeed in homogeneous nucleation experiments we observe that), at room temperature, the least stable polymorph, *β* glycine, is nucleated first and then transforms into the *α* or *γ* polymorph depending on the amount of water present. However, in the single particle experiments, this Ostwald-rule route is not always followed, and the *α* or *γ* polymorph could be seen to form directly from amorphous phase. There are two possible explanations: One is that the *β* polymorph was short lived (less than the 1-s integration time used for Raman spectroscopy) and could not be temporally resolved or that the electric field of the laser indeed affects the alignment of glycine molecules and the *α* or *γ* polymorph is directly formed.

There is mounting evidence for the existence of amorphous intermediates on-path or off-path from supersaturated solution to crystal. Similar to the case of sodium thiosulfate pentahydrate reported recently,[26] in aqueous glycine solutions the amorphous intermediate—detected here for the first time, using *in situ* low-frequency Raman spectroscopy—is metastable to such an extent that it inhibits crystal nucleation for more than 18 hours. However, it is also found here that the same amorphous intermediate is key to the process of laser induced nucleation in glycine solutions. The requirement for "aging" strongly suggests that a similar mechanism is at work in other examples of laser induced nucleation.

## ASSOCIATED CONTENT

The Supporting Information is available free of charge on the ACS Publications website at DOI:

Methods, kinetics of particle formation, temperature rise in the laser focus, and supplementary figures.

Supplementary Movie 1, laser induced nucleation and growth.

Supplementary Movie 2, transformation of particle into crystal.

## AUTHOR INFORMATION

Corresponding Author

Klaas Wynne, School of Chemistry, University of Glasgow, Glasgow G12 8QQ, U.K.; 0000-0002-5305-5940; Email: Klaas.wynne@glasgow.ac.uk

Authors

Zhiyu Liao, School of Chemistry, University of Glasgow, Glasgow G12 8QQ, U.K.; 0000-0002-6447-5510

**NOTES**

The authors declare no competing financial interest.

The data that support the findings of this study are available in Enlighten: Research Data Repository (University of Glasgow) with the identifier: @@@

**ACKNOWLEDGMENT**

We thank the Engineering and Physical Sciences Research Council (EPSRC) for support through grants EP/J004790/1 and EP/J014478/1. This work received funding from the European Research Council (ERC) under the European Union's Horizon 2020 research and innovation program (grant agreement No. 832703). We gratefully acknowledge one of the reviewers for extensive constructive comments that has greatly improved this article.

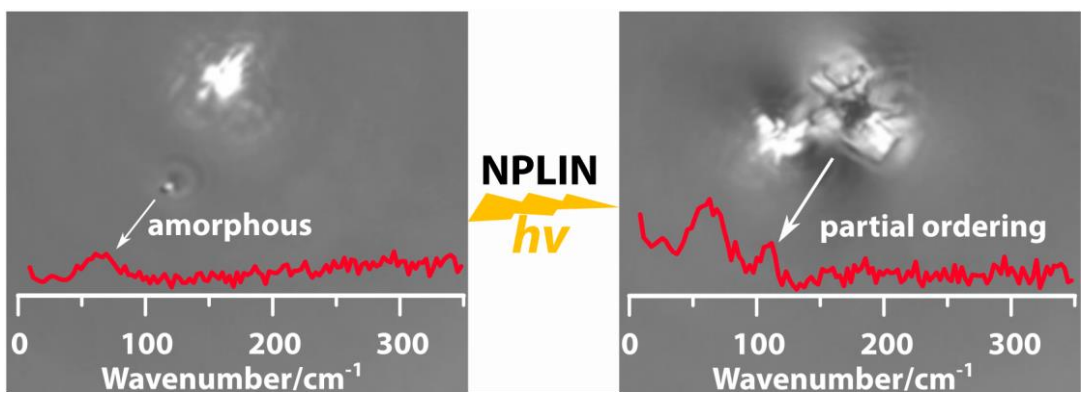

# Supporting Information – A metastable amorphous intermediate is responsible for laser-induced nucleation of glycine

Zhiyu Liao, Klaas Wynne

School of Chemistry, University of Glasgow, G12 8QQ, UK

## METHODS

**Sample preparation**. Glycine (PlusOne Glycine, GE Healthcare, 99.7%) and $D_2O$ (Sigma-Aldrich, 99.9%) were used as received without any further purification. Saturated glycine/$D_2O$ solutions were prepared by dissolving 0.22 g of glycine in 1 g of $D_2O$ in a clean glass vial. The solution was shaken at 60℃ for 8 hours at a speed of 500 rpm, then gradually cooled down to room temperature. For the samples used for microscopy and Raman spectroscopy, a droplet of 10 μL saturated solution was placed on the glass bottom of a Petri dish (12 mm Glass Based Dish, Thermo Scientific, which was ultrasonicated with HPLC water and dried in a nitrogen flow), along with 1 mL of the same solution on the side to slow evaporation. The dish was covered with a lid during the experiments.

**Microscopy and Raman setup**. The setup for microscopy and Raman spectroscopy was home-built on a double-deck inverted microscope (Olympus IX73), a schematic is depicted in 0. Phase contrast microscopy images and videos were captured by using a 60×/0.7 N.A. objective (Olympus, UCPlanFL N Ph2) and a CMOS camera (Teledyne Dalsa, Genie Nano-1GigE). Two laser sources were aligned into the microscope objective simultaneously: a high-power pulsed 1040 nm laser (Spectra-Physics, Spirit One, 8 W) or a high-power CW 1064 nm laser (BKtel photonics, MFL-1064-P400-FCAPC, 10 W) intended for optical tweezing and trapping, and a single-frequency 532 nm laser with linear polarization (Laser Quantum, gem 532, 500 mW) for Raman excitation. BragGrate[TM] bandpass and notch filters (OptiGrate) were used to achieve ultra-low frequency Raman spectroscopy, allowing detection of Raman scattering to frequencies as low as 10 $cm^{-1}$. For detection, a spectrometer (Andor, Shamrock 500i with 600 groove/mm grating) and a CCD camera (Andor, iDUS 401) were employed. An optical fiber with 50 μm core size was used for confocal Raman collection. The spectral resolution of the system is ~2 $cm^{-1}$.

**Dynamic Light Scattering (DLS)**. DLS measurements were performed using a particle size analyzer (Litesizer 500, Anton Paar) with laser light wavelength $\lambda$= 647 nm and accurate temperature control. All the intensity traces were recorded using back scattering ($\theta = 175^o$). The autocorrelation decay curves and mean hydrodynamic diameter of nano- and micro-particles was generated and estimated using the cumulant method provided by the Kalliope software.

**Data processing**. Raman spectra obtained from glycine/$D_2O$ solution, pure $D_2O$ and glycine micro-particles in solution in the low frequency region (10 - 500 $cm^{-1}$) were Bose-Einstein corrected to remove the temperature dependence of the vibrations. The processing was done in MATLAB (MathWorks) by applying the Bose-Einstein factor to the raw spectra $I(\omega)/(1+n(\omega))$ after background subtraction (dark counts of CCD detector), where $n(\omega) = (exp\textasciicircum(-\hbar\omega/k_B T) - 1)^{-1}$, $\omega$ is the angular frequency, and $I$ the Raman amplitude at $\omega$.

## TEMPERATURE RISE IN THE LASER FOCUS

The local temperature in the laser focus was measured by using anti-Stokes and Stokes Raman scattering in a glycine/$D_2O$ solution. 0(a) shows the intensity ratio of anti-Stokes and Stokes scattered light in the low-frequency region (elastic scattering) as a function of laser power. To maximize the signal-to-noise ratio, 1000 spectra with a 1-s integration time each were taken at each laser power and averaged. The temperatures are estimated by taking the ratio $I_{AS}/I_S$ = exp(-hν/$k_B$T), where $I_S$ and $I_{AS}$ are the Stokes and anti-Stokes intensities and ν the Raman shift. 0(b) shows that the local temperature increases ~3K when laser intensity increases from 20 mW to 50 mW. The room temperature in the lab is maintained at ~20℃.

## KINETICS OF PARTICLE FORMATION

To make sure the particles are not contamination in the form of dust, a kinetic study of particle formation in glycine solution was carried out. In the experiments described in the main text, the supersaturation of the solutions is typically increased in a controlled fashion by using additional droplets of solution (to slow down evaporation) or by using Drierite particles (to speed up evaporation). In order to get a handle on kinetics, dynamic light scattering (DLS) is an ideal technique. However, the controlled increase of supersaturation used is not applicable to DLS. Therefore, an alternative method was used to increase supersaturation.

A supersaturated glycine solution (S = 1.2) was prepared, filtrated using an Anotop 20-nm pore-size filter, and transferred to a cuvette. The 4.5 mL cuvette was filled with 1 mL solution, capped straightaway, and observed using DLS at room temperature for 15 hours. The time-domain DLS data exhibit a mono-exponential decay corresponding to a hydrodynamic diameter of ~ 1 nm consistent with glycine molecules (see 0(a) and (b)), demonstrating the absence of common room dust as well as the absence of particles at S = 1.2.

Next the temperature was raised to 50°C, which causes significant and controlled solvent evaporation (due to condensation near the top of the cuvette, see the photo in 0(e)). DLS measurements were taken every 3 hours over 18 hours at a constant temperature of 50℃. After a few hours at 50°C, an additional slow component appears in the DLS autocorrelation (see 0(c)) corresponding to particles with an average size rising from *circa* 150 nm to an average of *circa* 500 nm at longer waiting times. The noise in these DLS data is caused by the low concentration of the particles giving rise to considerable shot noise. Importantly, no crystals could be observed by eye after the measurements.

These results are consistent with previous studies describing the formation of mesoscale clusters (hundreds of nanometers in diameter) in supersaturated[1] and undersaturated[2] aqueous glycine solutions after dissolution, and in mutual equilibrium with molecular clusters (usually a few solute molecules).

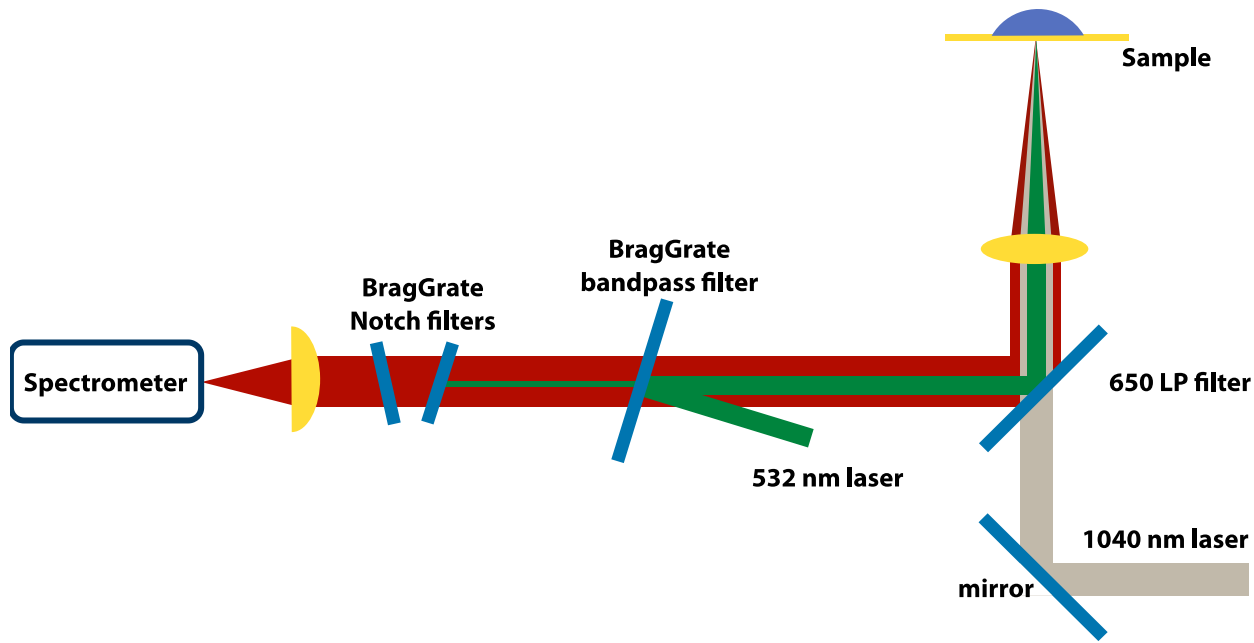

**Figure S1. Optical tweezing and broadband Raman spectroscopy system**. Schematic of the optical tweezing and broadband Raman spectroscopy system based on an inverted double-deck microscope.

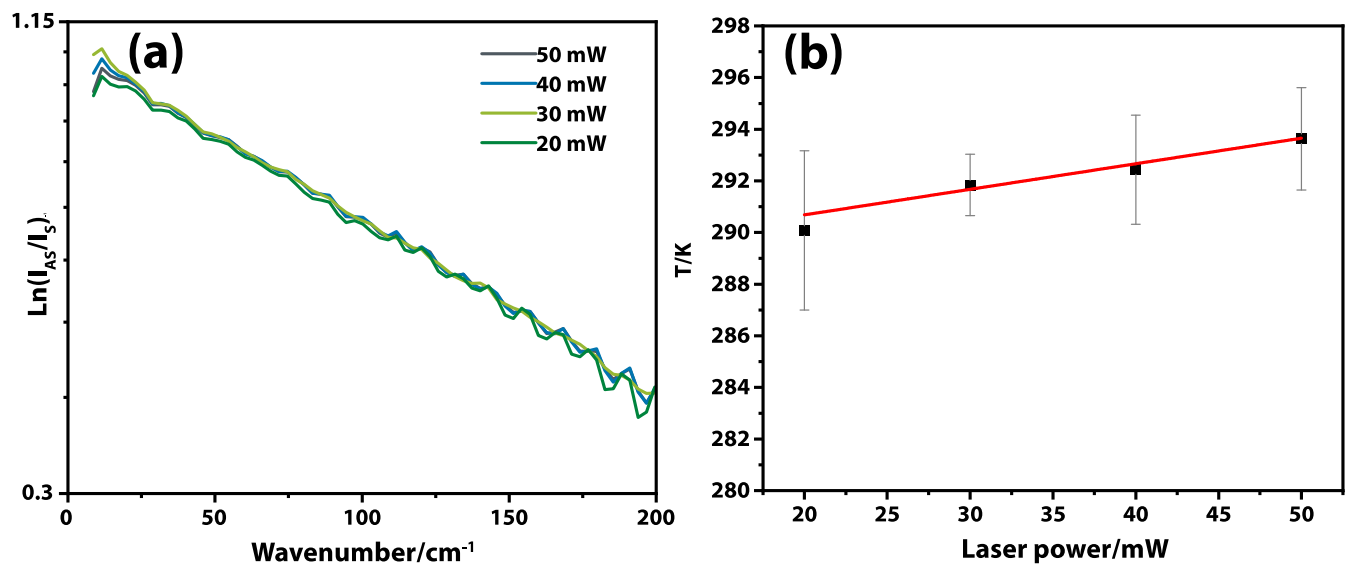

**Figure S2. Temperature estimation using anti-Stokes and Stokes Raman scattering**. (a) Intensity ratio of anti-Stokes and Stokes Raman scattering of glycine/D2O solution in the low-frequency region on a natural logarithmic scale. (b) Estimated temperate in the laser focus as function of laser power.

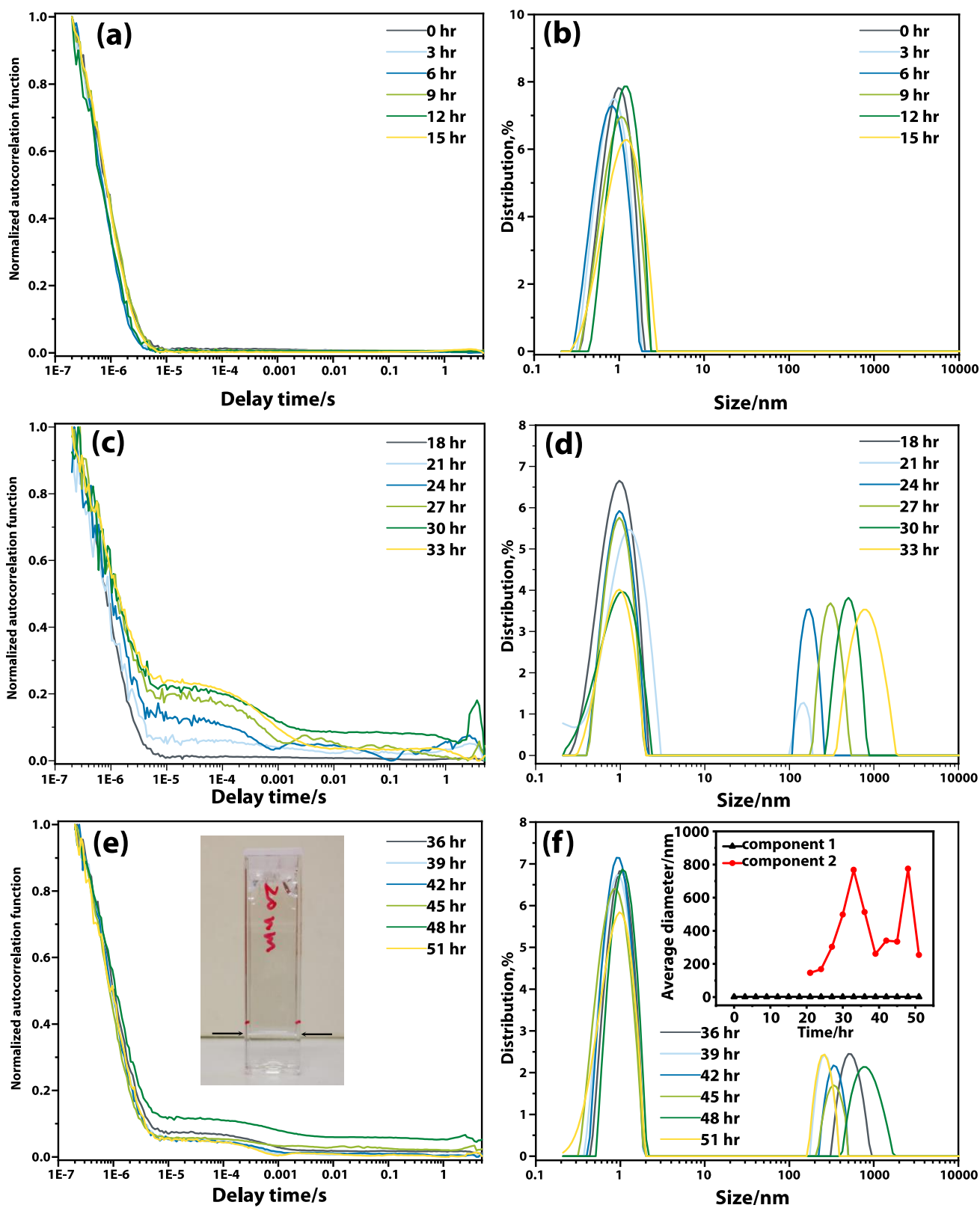

**Figure S3. DLS investigation of particles formation in glycine/D2O solution after filtration.** (a) Autocorrelation decay curves obtained at room temperature for the first 15 hours. (b) Corresponding size distribution. (c) Autocorrelation decay curves obtained from the same sample kept at 50°C for an additional 18 hours. (d) Corresponding size distribution. (e) Autocorrelation decay curves obtained from the same sample kept at 50°C for another 18 hours, inset picture shows the solution level before (indicated by red dots) and after (indicated by black arrows) measurements. (f) Corresponding size distribution. The inset shows the temporal evolution of the average diameter of the particles over time.

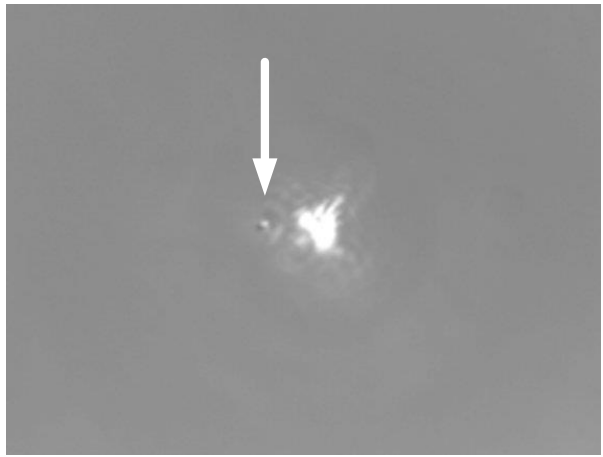 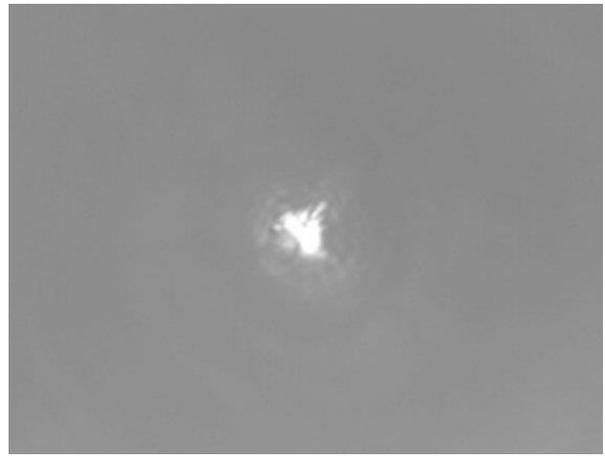

t = 8.000 s     t = 8.733 s

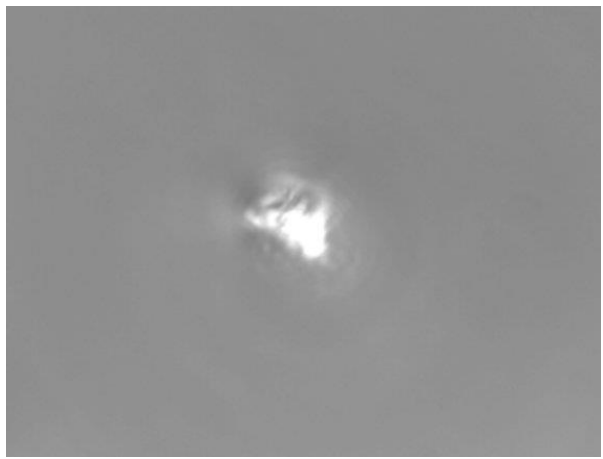 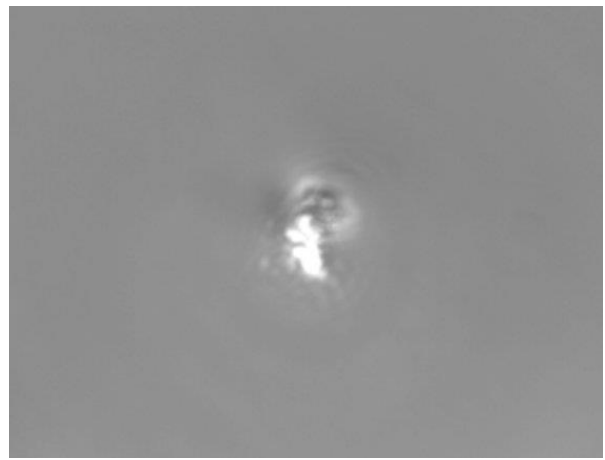

t = 8.766 s     t = 8.800 s

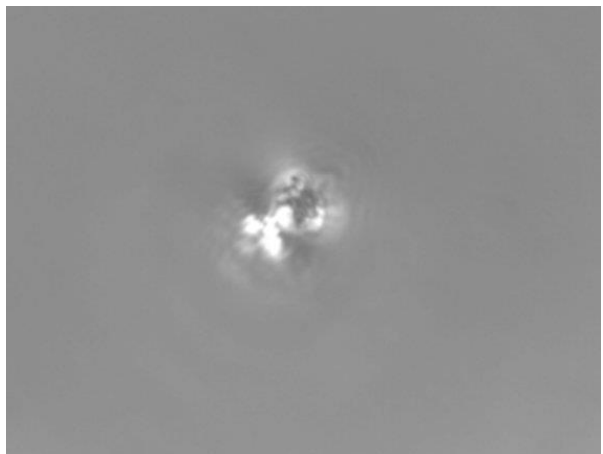 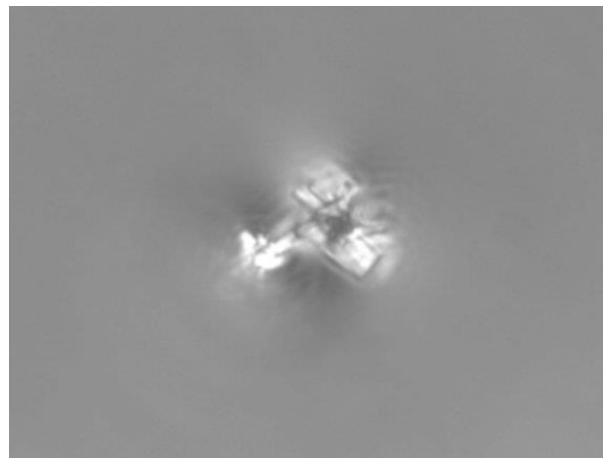

t = 8.866 s     t = 9.066 s

**Figure S4. Selected frames corresponding to the experiment shown in Figure 3**. The particle (white arrow) is brought into laser focus at *t* = 8.733 s, and nucleation is instantly induced by the high-power laser irradiation (380 mW), mesoscale clusters form in the next frame at *t* =8.766 s, and rapidly grow in the next few frames.

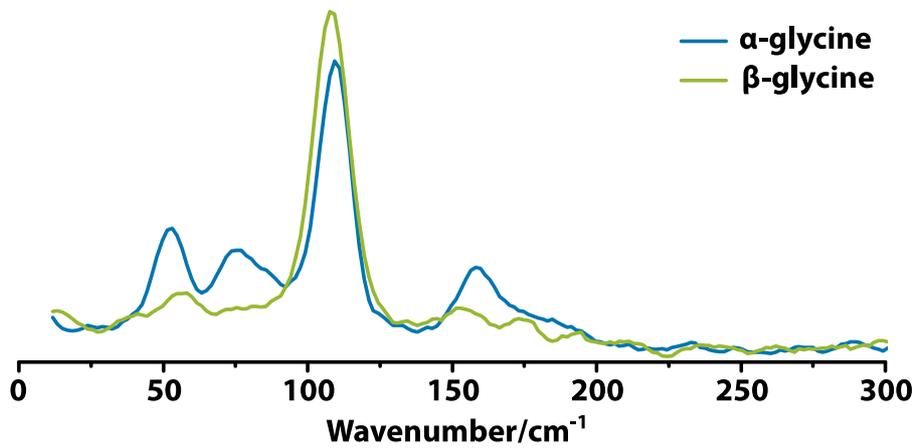

**Figure S5. Low-frequency Raman spectra of different glycine polymorphs.** Low-frequency Raman spectra obtained from a crystal with a needle-like habit corresponding to β glycine (green) and one with a prismatic habit corresponding to α glycine (blue) shown in Figure 3(f).

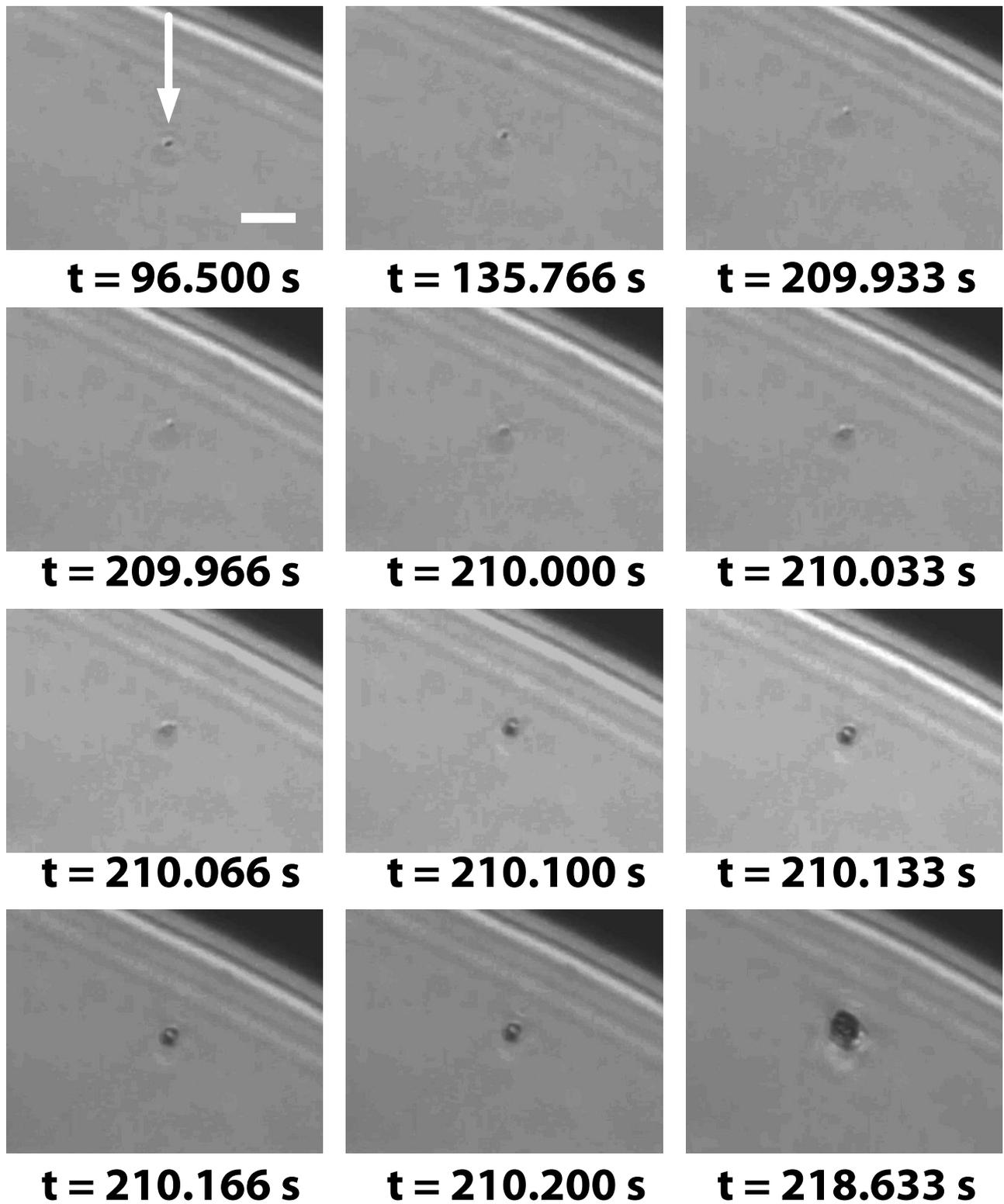

**Figure S6. Selected frames corresponding to the experiment shown in Figure 4.** The contrast has been increased to show the particle (white arrow) and the mesoscale cluster in greater detail. The particle is trapped by the laser at $t = 22$s and remains unchanged up to $t = 209.966$s. At this point, the particle begins to expand and then changes shape to form a mesoscale cluster in just a few frames. After about 8s, the mesoscale cluster has grown to a few micrometers in diameter. Scale bar is 5 μm.

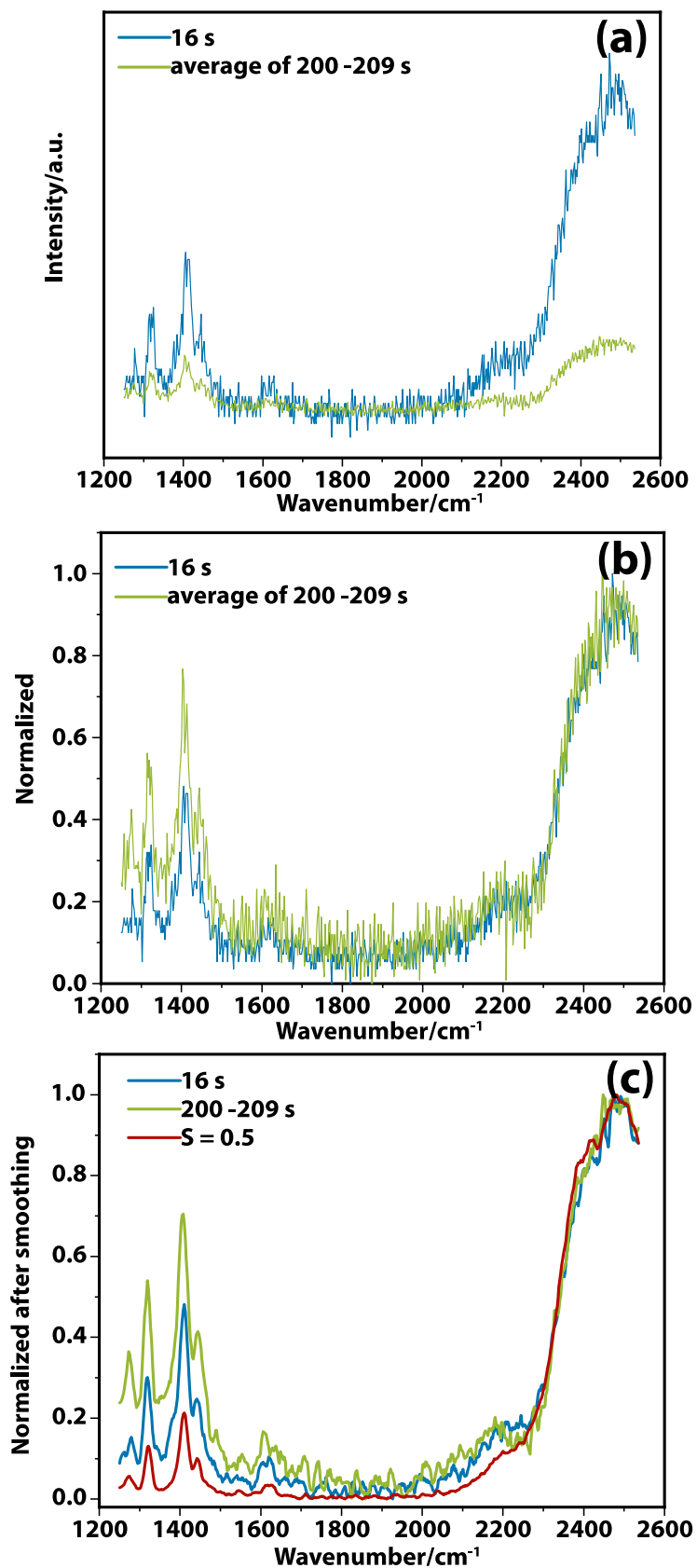

**Figure S7.** Comparison of Raman spectra with (200 -209 s) and without (16 s) particle trapped in laser focus in the example of Figure 4 shown in the main text. (a) Raw spectra. (b) Normalization using the O-D band of $D_2O$. (c) Normalized spectra after smoothing and comparison with the starting solution (S = 0.5) before any laser irradiation.

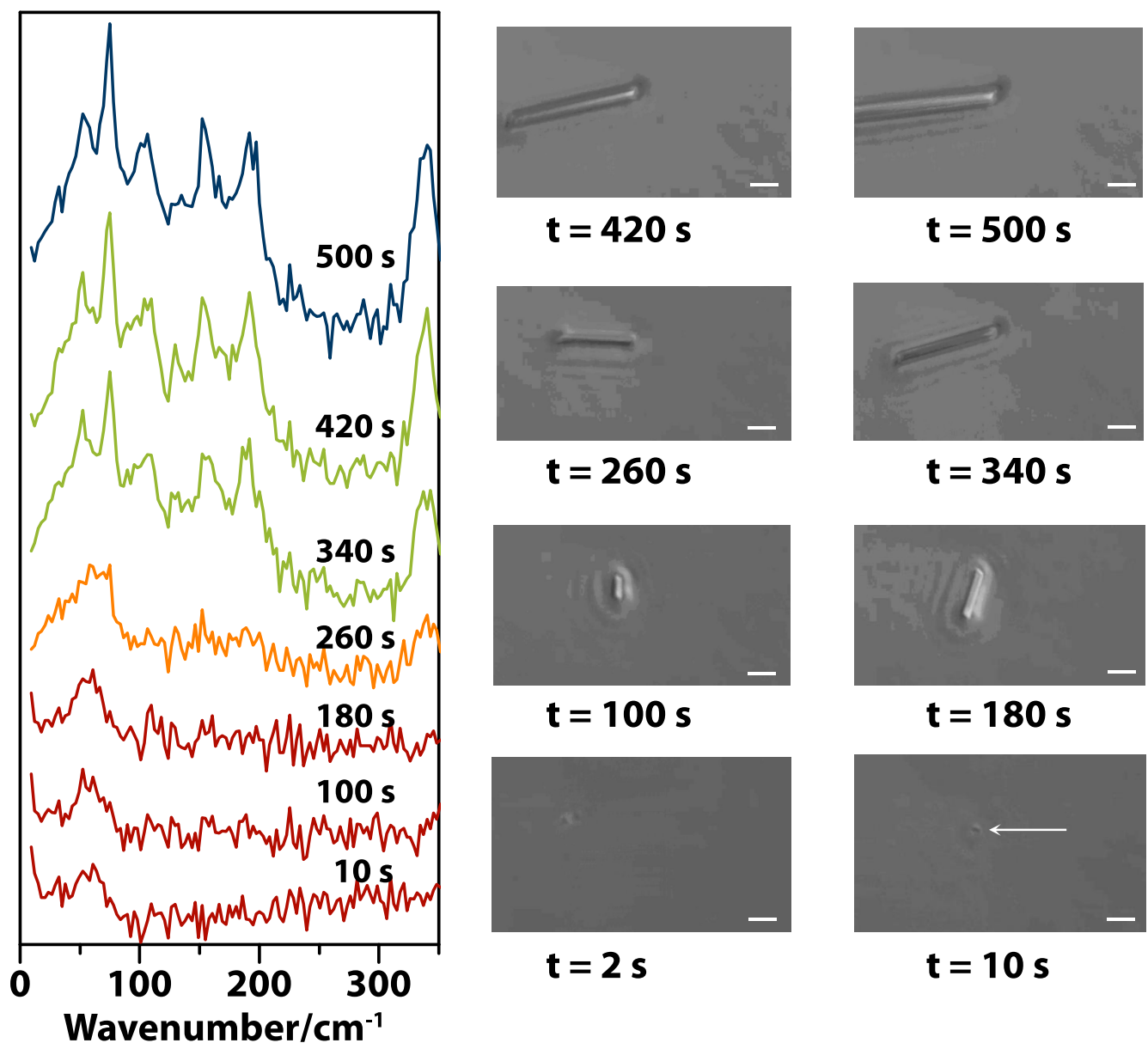

**Figure S8. Additional example of laser-induced nucleation of a glycine particle in solution**. Another example of spectral evolution due to intermediate stages between microscopic particle and a fully-fledged crystal, time-matched microscopy images shown on the right. The spectra are subtracted by the spectrum of $D_2O$ taken at t = 2 s when the particle was not present in the laser focus.

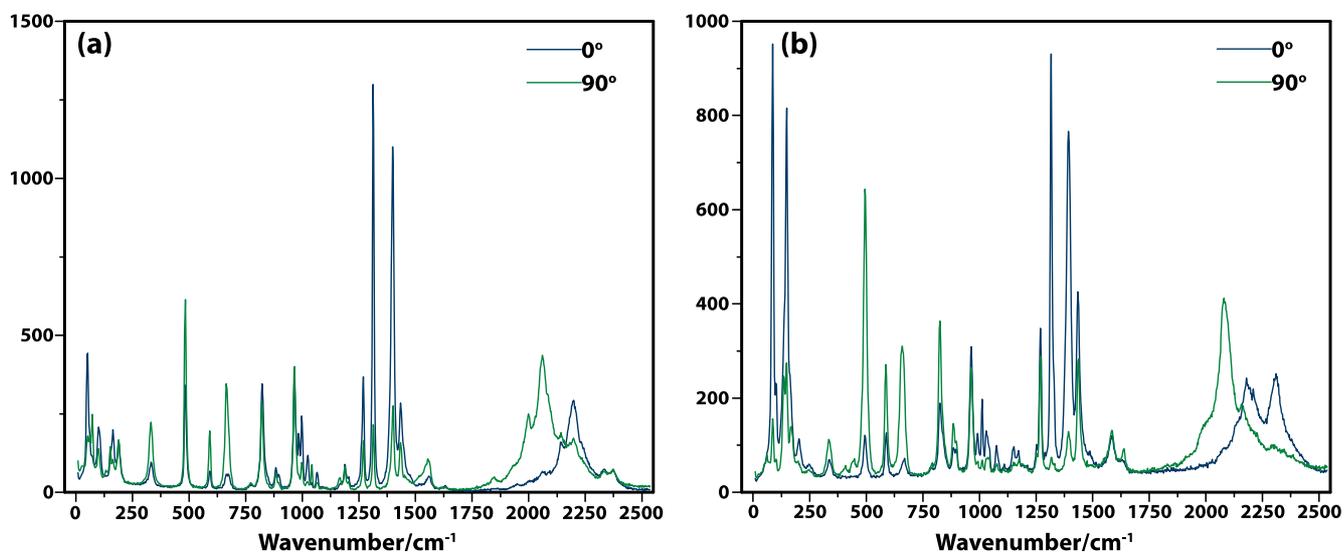

**Figure S9. Polarized Raman spectroscopy of different glycine polymorphs**. Polarized Raman spectra of (a) α-glycine and (b) γ-glycine crystals.

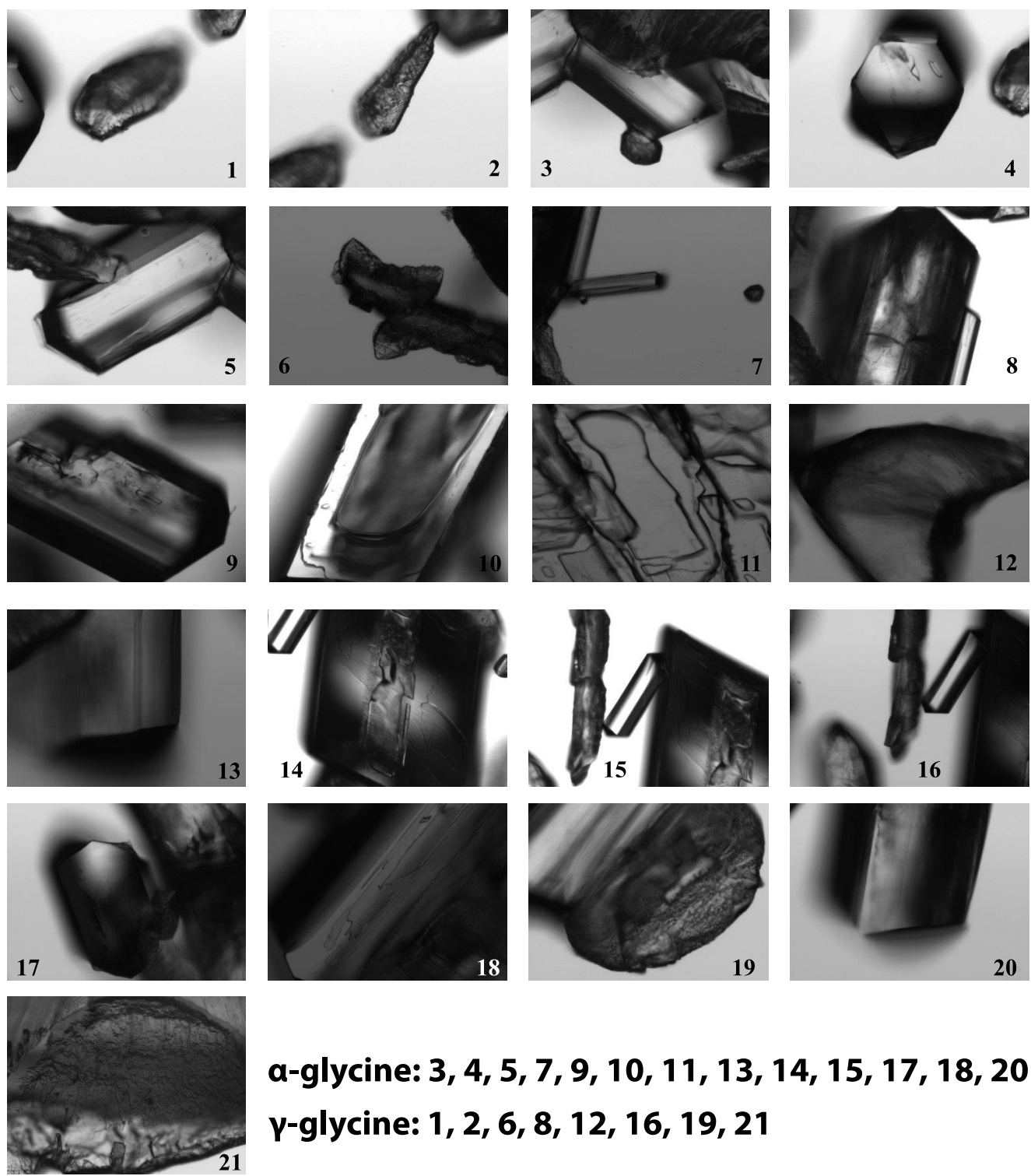

**Figure S10. Microscopy images of α- and γ-glycine formed in D₂O solution after laser irradiation and subsequent drying.** Raman spectra confirm that all the crystals with polished surface are α-glycine, while crystals with rough surface are γ-glycine.

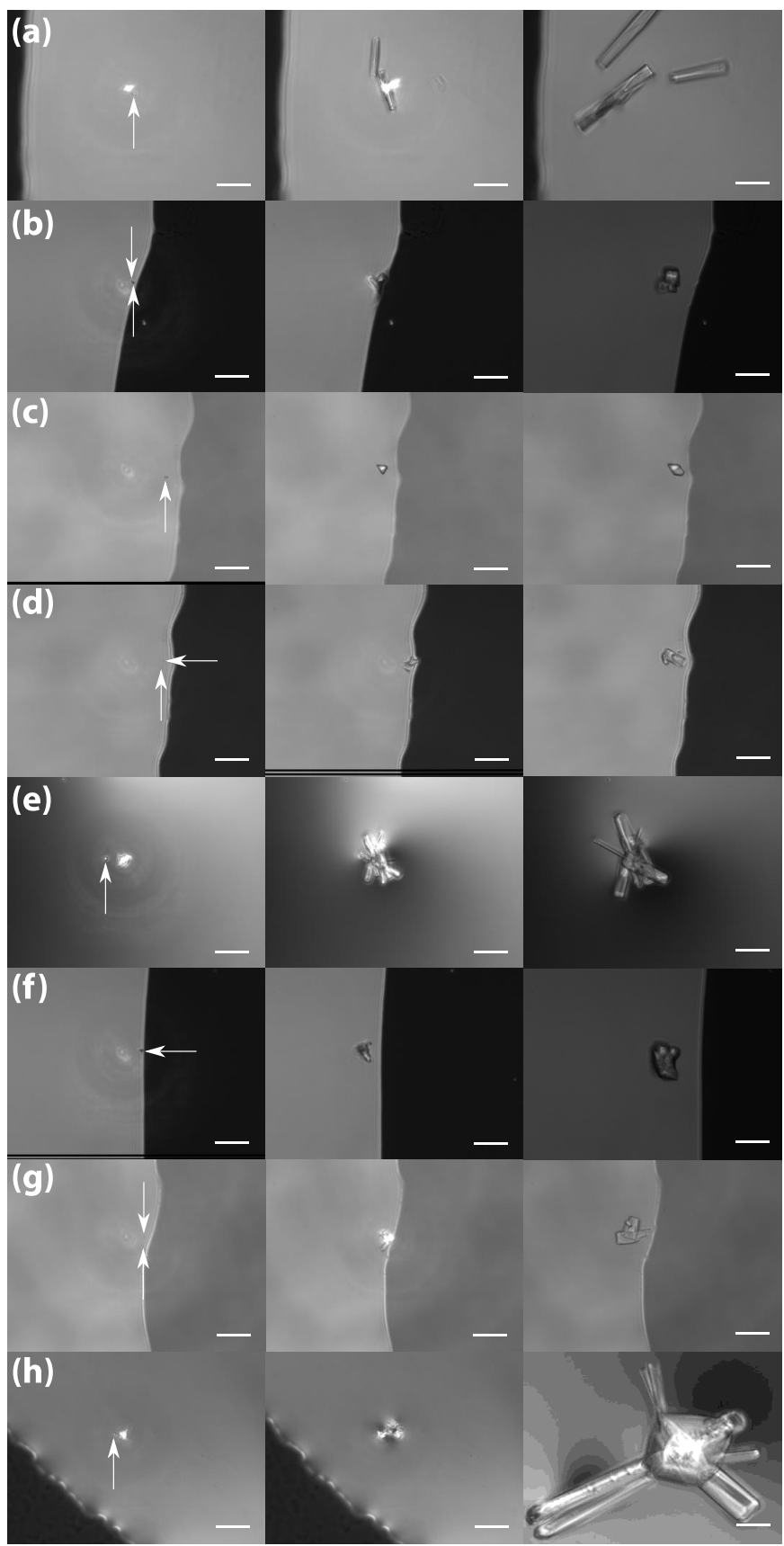

**Figure S11.** Microscopy images of 8 glycine particles (a-h) right before (left) and after (middle) laser-induced nucleation using 1040 nm at 380 mW, final crystals shown on the right. All scale bars, 30 μm. Arrows indicate the glycine particles. Panels (b), (d) and (g) show two particles simultaneously exposed to laser irradiation and subsequently generate two crystals. Panel (e) is an example where one particle results in 6 crystals.

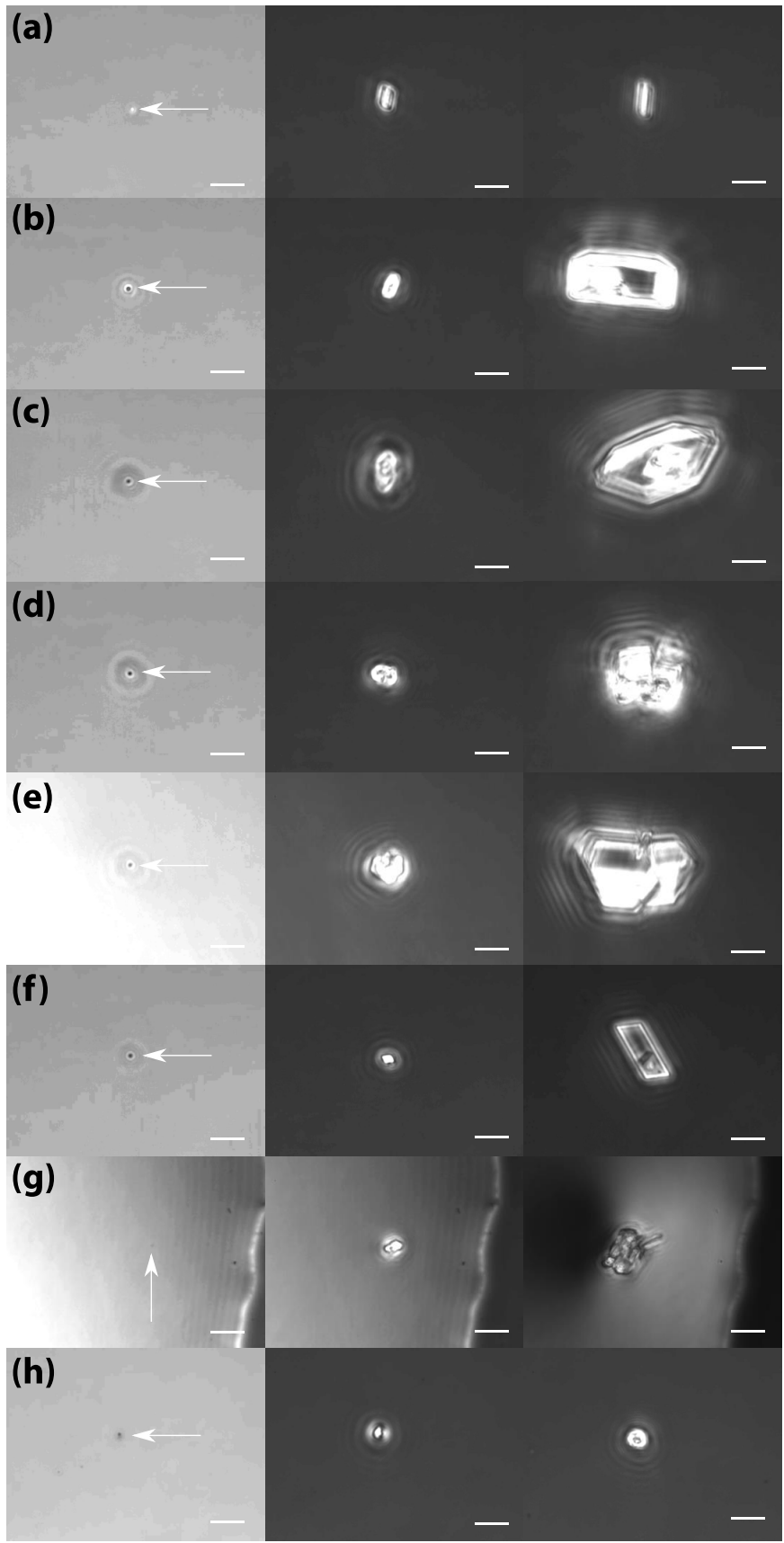

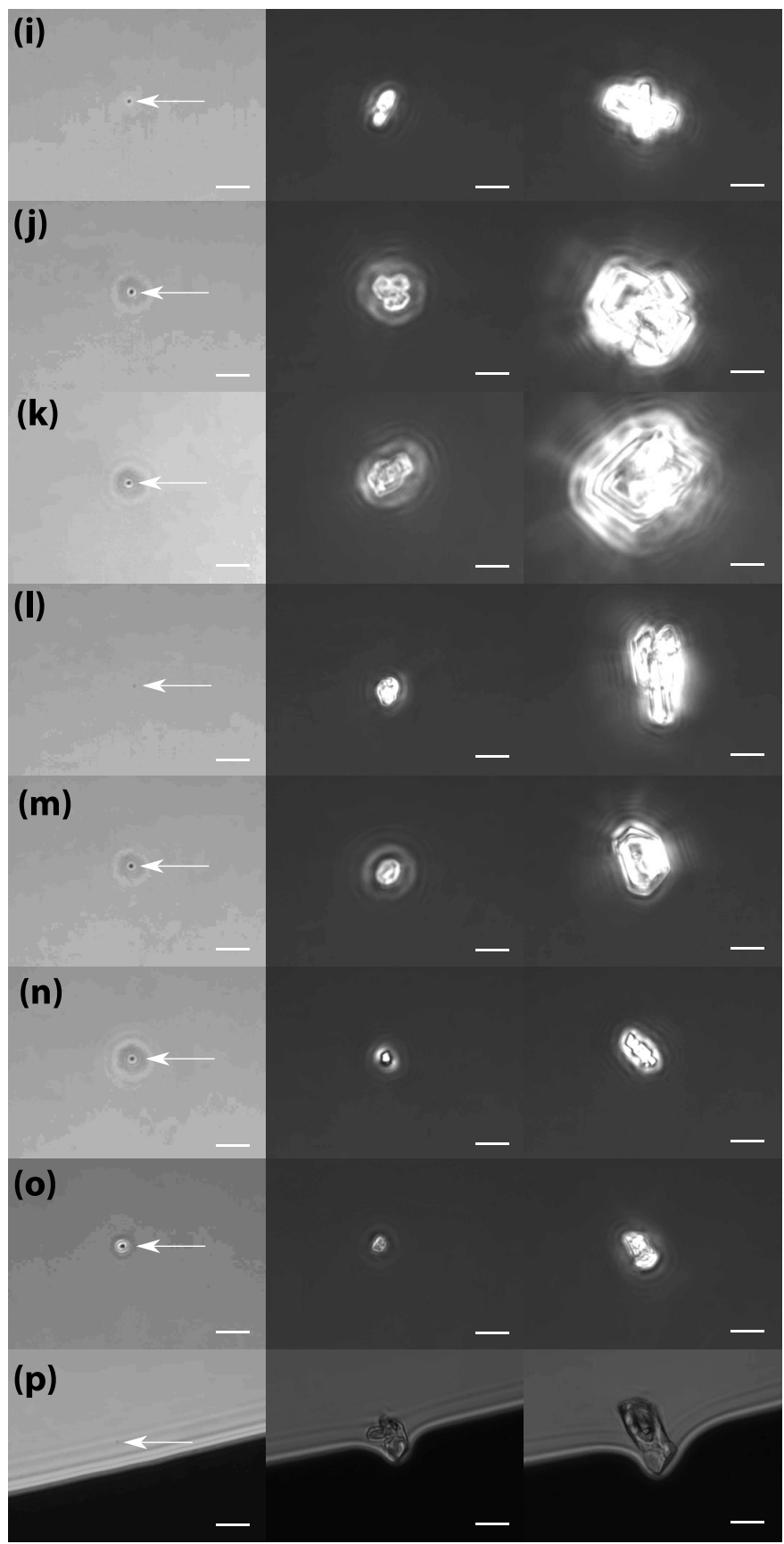

**Figure S12.** Microscopy images of 16 glycine particles (a-p) right before (left) and after (middle) laser-induced nucleation using 532 nm at 50 mW, final crystals shown on the right. All scale bars, 10 μm. Arrows indicate the glycine particles. Brightness of the images on the left are enhanced to improve the contrast.